\documentclass[10pt]{IEEEtran}
\IEEEoverridecommandlockouts
\usepackage{cite}
\usepackage{amsmath,amssymb,amsfonts}
\usepackage{graphicx}
\usepackage{textcomp}
\usepackage{amsmath}
\usepackage{booktabs} 
\usepackage{multirow}
\usepackage{subcaption}
\usepackage{hyperref}
\usepackage{tabularx}
\usepackage{url}            
\usepackage{booktabs}       
\usepackage{amsfonts}       
\usepackage{nicefrac}       
\usepackage{microtype}      
\usepackage{xcolor}         
\usepackage{algorithm}
\usepackage{colortbl}
\usepackage{algpseudocode}
\usepackage{amsmath}
\usepackage{tikz}
\usepackage{orcidlink}
\usepackage[skip=0pt,font=footnotesize]{caption}
\usepackage{float}
\usepackage{stfloats}

\usepackage{orcidlink}
\usepackage{comment}
\usepackage{amssymb}
\usepackage{graphicx}
\usepackage{braket}
\definecolor{headerblue}{RGB}{214,226,240}
\definecolor{rowgray}{RGB}{245,247,250}
\definecolor{highlight}{RGB}{231,242,255}
\usepackage{setspace}
\usepackage{xcolor}
\def\BibTeX{{\rm B\kern-.05em{\sc i\kern-.025em b}\kern-.08em
    T\kern-.1667em\lower.7ex\hbox{E}\kern-.125emX}}
    \usepackage{setspace}
\begin{document}

\title{Towards Fair Benchmarking of Quantum Transfer Learning for Visual Classification}


\author{\IEEEauthorblockN{Nouhaila Innan\textsuperscript{1,2}, Saim Rehman\textsuperscript{1,2}, and Muhammad Shafique\textsuperscript{1,2}\\
\IEEEauthorblockA{
\textsuperscript{1}eBRAIN Lab, Division of Engineering, New York University Abu Dhabi (NYUAD), Abu Dhabi, UAE\\
\textsuperscript{2}Center for Quantum and Topological Systems (CQTS), NYUAD Research Institute, NYUAD, Abu Dhabi, UAE\\
\{nouhaila.innan, sr7849, muhammad.shafique\}@nyu.edu\\
}}}

\maketitle
\thispagestyle{empty}
\pagestyle{empty}
\begin{abstract}
Quantum Transfer Learning (QTL) offers a promising approach for visual quantum machine learning under near-term constraints, where limited qubit counts, shallow circuit depths, and costly hybrid optimization restrict end-to-end quantum training. In this setting, pretrained classical backbones can extract high-level visual features, while compact quantum modules operate as trainable classification heads. However, existing QTL results are difficult to compare because they often differ in datasets, preprocessing, backbone settings, qubit budgets, circuit designs, optimization choices, and reporting protocols. This work presents a controlled benchmarking methodology for evaluating representative QTL methods under a unified transfer-learning pipeline. The benchmark compares DQN-QTL, QPIE-QTL, AE-CQTL, PVCQTL, and ED-QTL under shared preprocessing rules, frozen-backbone settings, training conditions, and reporting metrics. The evaluation focuses on Fashion-MNIST and Hymenoptera Ants vs. Bees as the two main datasets, while CIFAR-10 is used to provide additional configuration-level evidence on a harder natural-image task. Beyond predictive performance, the benchmark analyzes circuit size, trainable parameters, quantum parameters, training time, and architectural sensitivity to qubit count and circuit depth. The results show that no single QTL family dominates across all settings: performance depends on the dataset, encoding strategy, circuit design, and computational cost. These findings highlight the need for resource-aware QTL evaluation and provide guidance for selecting hybrid quantum-classical transfer models under near-term resource constraints.
\end{abstract}

\begin{IEEEkeywords}
Quantum Transfer Learning, Image Classification, Quantum Machine Learning 
\end{IEEEkeywords}

\section{Introduction}
Quantum machine learning (QML) has become an active research direction for exploring how quantum models can support learning tasks in areas such as image classification, healthcare, finance, and pattern recognition~\cite{biamonte2017quantum,innan2025lep,choudhary2025hqnn,siddiqui2025quantum,innan2025qnn,ahmad2026quantum,kashif2026design,peral2024systematic}. Many near-term QML approaches rely on hybrid quantum-classical models, where parameterized quantum circuits (PQCs) are integrated with classical learning components~\cite{innan2025next}. However, current QML models remain constrained by limited qubit counts, shallow circuit depths, noisy quantum operations, and costly hybrid optimization~\cite{vyskubov2026scaling,njiki2026robustness}. These constraints make end-to-end quantum training difficult for high-dimensional data such as images. As a result, quantum circuits are often used as compact trainable modules within larger classical pipelines rather than as complete standalone models.

This setting naturally motivates quantum transfer learning (QTL) \cite{mari2020transfer,azevedo2022quantum}. Instead of training a full quantum model directly on raw images, a pretrained classical backbone can be used to extract high-level visual features, while a quantum circuit serves as a classifier or representation-learning head. This design is well aligned with near-term QML because the classical backbone reduces input dimensionality and provides informative feature representations, while the quantum component can operate within limited qubit and circuit-depth budgets. QTL therefore provides a structured way to study hybrid quantum-classical learning under realistic resource constraints.

Several QTL approaches have been proposed in recent years \cite{mari2020transfer,buonaiuto2024quantum,kim2023classical,mogalapalli2022classical,qi2022classical,otgonbaatar2023quantum}, including dressed quantum transfer heads, probability-inspired encoding strategies, amplitude-encoded quantum models, post-variational quantum transfer designs, and distillation-based hybrid methods. These approaches differ in how classical features are reduced, how information is encoded into quantum states, how the PQC is structured, and how quantum measurements are converted into final predictions. However, despite these developments, it remains difficult to draw clear conclusions about which QTL strategies are more effective or efficient. Reported results often rely on different datasets, pretrained backbones, input resolutions, qubit counts, circuit depths, optimization settings, and evaluation protocols.

\textit{This lack of controlled comparison limits the interpretation of QTL results.} 
A performance improvement may come from the transfer strategy itself, but it may also result from a stronger classical backbone, a larger number of qubits, deeper circuits, different preprocessing choices, or longer training. Similarly, accuracy alone does not indicate whether a QTL method is practical, since two models with similar predictive performance may differ substantially in parameter count, quantum circuit size, and training time. For near-term QTL, these cost factors are central to understanding whether a method is useful beyond a single reported result.

\textit{To address this issue, this work introduces a controlled benchmarking methodology for evaluating representative QTL methods under shared datasets, model settings, and reporting criteria.} 
The benchmark compares several QTL families within a common transfer-learning pipeline based on frozen pretrained feature extraction, quantum-compatible feature reduction, parameterized quantum processing, and final classification. The evaluation considers predictive performance, circuit size, trainable parameters, quantum parameters, training time, and architectural sensitivity to qubit count and circuit depth.

\textbf{The main contributions of this work are summarized as follows:}
\begin{itemize}
    \item We introduce a controlled QTL benchmarking methodology that evaluates representative quantum transfer learning families under shared preprocessing rules, frozen-backbone settings, training conditions, and reporting metrics.

    \item We compare representative techniques including DQN-QTL, QPIE-QTL, AE-CQTL, PVCQTL, and ED-QTL, covering dressed quantum transfer heads, multi-axis encoding, amplitude-encoded transfer, post-variational observable-based transfer, and teacher-guided classical-to-quantum knowledge transfer.

\item We evaluate diverse QTL methods on two main benchmark datasets, Fashion-MNIST and Hymenoptera Ants vs. Bees, and include CIFAR-10 as an additional evaluation setting for selected configurations.

    \item We analyze QTL performance beyond accuracy by jointly reporting predictive metrics, quantum circuit size, trainable parameters, quantum parameters, and training time.

    \item We study architectural sensitivity by examining the effect of qubit count and circuit depth, showing when larger or deeper quantum circuits improve performance and when they provide limited gains.

    \item We provide guidelines for selecting appropriate QTL configurations under near-term resource constraints by identifying performance-cost trade-offs across datasets and model families.
\end{itemize}

\section{Background}
\label{sec:background}

\subsection{Hybrid Quantum-Classical Models}

\begin{figure}[htpb]
    \centering
    \includegraphics[width=1\linewidth]{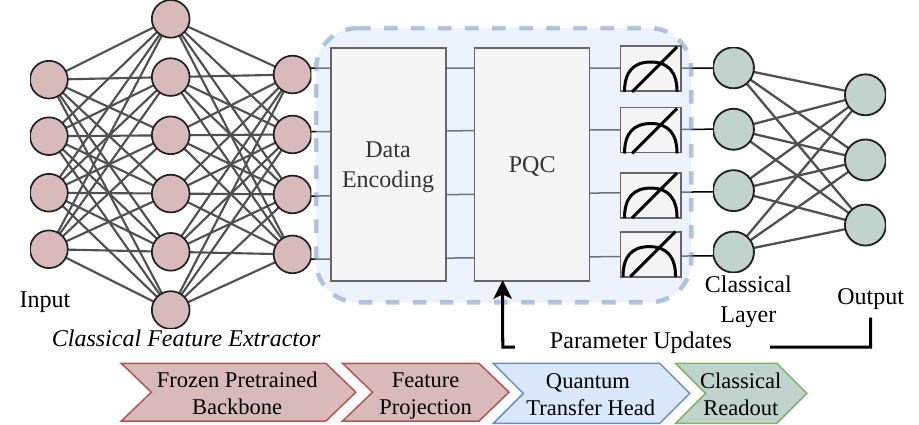}
\caption{Hybrid quantum-classical learning architecture and its QTL specialization. The general model uses classical layers before and after the quantum module for feature processing and prediction. In the QTL pipeline, a frozen pretrained backbone is followed by feature projection, a quantum transfer head, and a classical readout.}\label{fig:qtl_architecture}
\end{figure}

A hybrid quantum-classical model combines classical preprocessing, a parameterized quantum circuit (PQC), and a classical readout layer, as shown in Fig.~\ref{fig:qtl_architecture}. For an input feature vector \(\mathbf{x}\), the circuit prepares
$
|\psi(\mathbf{x};\boldsymbol{\theta})\rangle
=
U_{\boldsymbol{\theta}}U_{\mathrm{enc}}(\mathbf{x})
|0\rangle^{\otimes n},
$
where \(U_{\mathrm{enc}}(\mathbf{x})\) is the data-encoding unitary, \(U_{\boldsymbol{\theta}}\) is the trainable circuit with parameters \(\boldsymbol{\theta}\), and \(n\) is the number of qubits. Measuring observables \(\{M_j\}_{j=1}^{m}\) gives
$
z_j
=
\langle\psi(\mathbf{x};\boldsymbol{\theta})|
M_j
|\psi(\mathbf{x};\boldsymbol{\theta})\rangle, $ for $ j=1,\ldots,m,
$
where \(M_j\) is the \(j\)-th observable, \(m\) is the number of measurements, and \(z_j\) is its expectation value. The measurement vector \(\mathbf{z}=[z_1,\ldots,z_m]\) is passed to a classical readout for prediction.
\subsection{Quantum Transfer Learning}

In transfer learning~\cite{torrey2010transfer,weiss2016survey,zhuang2020comprehensive,plested2026deep}, a pretrained backbone \(\Phi(\cdot)\) maps input \(\mathbf{x}\) to a feature representation
$
\mathbf{r}=\Phi(\mathbf{x}),
$
where \(\mathbf{r}\) is then passed to a task-specific classifier. The backbone may be frozen or fine-tuned.

Quantum transfer learning replaces the conventional classifier with a quantum or hybrid head. A reduction map \(g_{\boldsymbol{\omega}}\), with trainable parameters \(\boldsymbol{\omega}\), converts \(\mathbf{r}\) into a quantum-compatible vector
$
\mathbf{v}
=
g_{\boldsymbol{\omega}}\!\left(\Phi(\mathbf{x})\right).
$
The vector \(\mathbf{v}\) is encoded and processed by the PQC,
$
|\psi(\mathbf{v};\boldsymbol{\theta})\rangle
=
U_{\boldsymbol{\theta}}U_{\mathrm{enc}}(\mathbf{v})
|0\rangle^{\otimes n},
$
and the resulting measurements \(\mathbf{z}\) are mapped to the prediction
$
\hat{y}=h_{\boldsymbol{\phi}}(\mathbf{z}),
$
where \(h_{\boldsymbol{\phi}}\) is the classical readout with parameters \(\boldsymbol{\phi}\), and \(\hat{y}\) is the predicted output. Thus, the quantum component lies in the transfer head, which encodes, transforms, and measures pretrained classical features for classification.

\section{Benchmark Design and Evaluation Protocol}
\label{sec:benchmark_protocol}

\begin{figure*}[htpb]
\centering
\includegraphics[width=\linewidth]{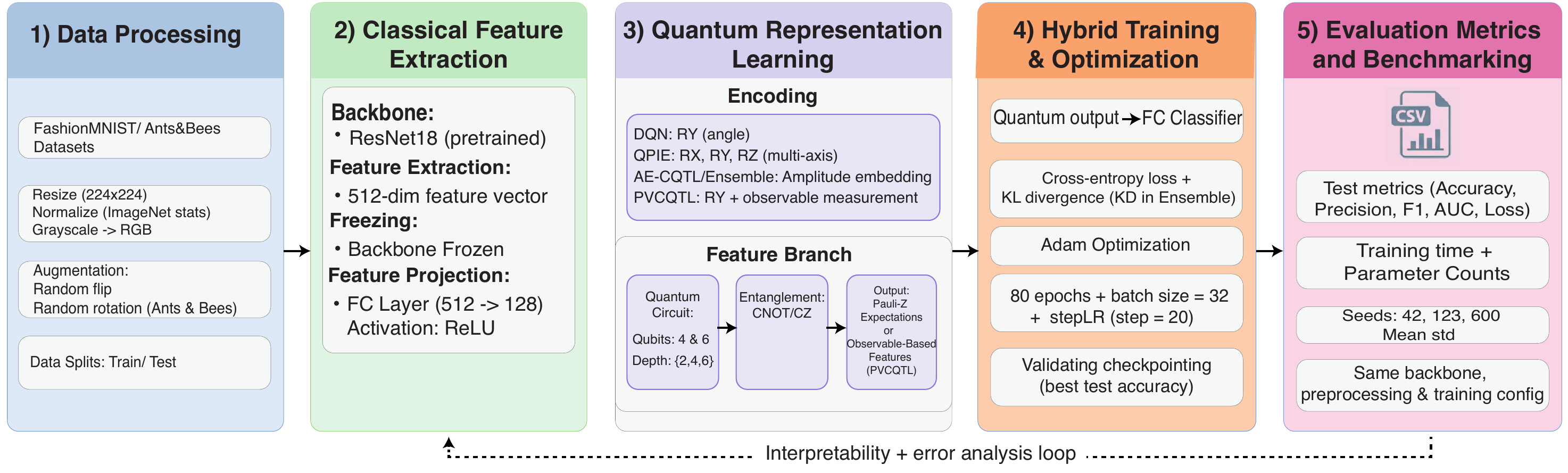}
\caption{Controlled protocol for evaluating QTL methods. Data processing, frozen-backbone feature extraction, hybrid training, and evaluation are standardized, while encoding, circuit architecture, and measurement design vary across methods.}
\label{meth}
\end{figure*}

As illustrated in Fig.~\ref{meth}, the benchmark compares DQN-QTL, QPIE-QTL, AE-CQTL, PVCQTL, and ED-QTL under a shared transfer-learning protocol. The comparison considers predictive performance, computational cost, and sensitivity to qubit count and circuit depth.

\subsection{Benchmark Objectives and Scope}

The benchmark examines three questions: how QTL families perform under the same pipeline, what computational and architectural costs they require, and how quantum design choices affect performance. This supports a cost-aware comparison rather than an accuracy-only ranking.

\subsection{Datasets and Shared Pipeline}

Fashion-MNIST and Hymenoptera Ants vs.\ Bees are the main datasets, representing structured grayscale classification and low-data natural-image transfer, respectively. CIFAR-10 is evaluated only for selected configurations and is not used for full cross-family ranking.

All methods use a frozen pretrained ResNet18 backbone. Extracted features are reduced to a quantum-compatible representation, processed by a parameterized quantum circuit, and passed to a classical readout. This setup isolates differences in feature reduction, encoding, circuit structure, and measurement design.

\subsection{Controlled Evaluation}

Input resizing, normalization, dataset splits, and class balance are standardized. Most configurations use \(4\) or \(6\) qubits, while AE-CQTL uses \(9\) qubits to amplitude encode the \(512\)-dimensional backbone output. Circuit depth is fixed where possible in the main comparison and varied separately in the scaling study.

Performance is evaluated using accuracy, precision, recall, F1-score, and ROC-AUC. Cost is measured through trainable parameters, quantum parameters, circuit size, and training time. Results are reported as mean \(\pm\) standard deviation across repeated runs, with reduced datasets or training budgets explicitly identified.

\section{Quantum Transfer Learning Techniques Compared}
\label{sec:qtl_methods}

This section describes the QTL techniques evaluated in the benchmark. All methods follow the shared transfer-learning pipeline defined in Section~\ref{sec:benchmark_protocol}: a pretrained classical backbone extracts image features, a classical reduction stage maps these features to a quantum-compatible representation, a quantum module processes the reduced features, and a classical readout produces the final prediction. The methods mainly differ in their feature reduction strategy, quantum encoding, parameterized circuit design, and measurement/readout mechanism.

For consistency, the evaluated methods are referred to as Dressed Quantum Network QTL (DQN-QTL), Quantum Parallel Information Exchange-inspired Quantum Transfer Learning (QPIE-QTL), Amplitude-Encoding Classical-Quantum Transfer Learning (AE-CQTL), Post-Variational Classical-Quantum Transfer Learning (PVCQTL), and Ensemble-Distillation Quantum Transfer Learning (ED-QTL).
\subsection{DQN-QTL}
DQN-QTL is used as the QTL baseline in this benchmark. The method follows the idea of placing a trainable quantum circuit between classical layers, forming a hybrid head on top of a pretrained feature extractor \cite{bhowmik2025quantum}. In this setting, the classical backbone provides a transferred representation, while the quantum circuit acts as a compact trainable module for classification.
In our implementation, the frozen ResNet18 backbone extracts a $512$-dimensional feature vector from each image. This vector is passed through a classical pre-network with hidden dimension $128$, then reduced to match the number of qubits used by the quantum circuit. The reduced features are scaled using a $\tanh$ activation and encoded into the quantum circuit using $R_Y$ angle encoding. The quantum circuit applies repeated layers of trainable $R_X$, $R_Y$, and $R_Z$ rotations followed by cyclic CNOT entanglement. Pauli-$Z$ expectation values are measured from all qubits and passed to a final classical linear classifier.
Within the benchmark, dressed QTL represents a direct transfer setting, where the pretrained classical representation is mapped into a compact angle-encoded quantum classifier.

\subsection{QPIE-Inspired QTL}

The QPIE-inspired QTL method is included to evaluate a richer quantum input encoding strategy. Rather than reproducing the full original QPIE architecture \cite{guo2025quantum}, this benchmark adopts the central idea of distributing classical information across multiple rotation channels before quantum processing. Compared with single-axis angle encoding, this design assigns multiple input rotations to each qubit, allowing the quantum circuit to receive a more expressive feature representation.

In our benchmark implementation, the frozen ResNet18 backbone extracts a $512$-dimensional feature vector. A classical reducer network with hidden dimension $128$ maps this feature vector to $3n$ values, where $n$ is the number of qubits. These values are reshaped so that each qubit receives three input angles corresponding to $R_X$, $R_Y$, and $R_Z$ rotations. After this multi-axis input encoding, the quantum circuit applies repeated variational layers composed of trainable $R_X$, $R_Y$, and $R_Z$ rotations, followed by cyclic CNOT entanglement. The circuit output is obtained through Pauli-$Z$ expectation values, which are passed to a final classical linear classifier.

Within the benchmark, QPIE-inspired QTL represents an encoding-focused transfer strategy. Its main difference from dressed QTL is that it encodes each reduced feature representation through multi-axis rotations rather than a single $R_Y$ encoding path.

\subsection{AE-CQTL}

AE-CQTL is an amplitude-encoding-based classical-quantum transfer learning method designed to process pretrained feature representations through a quantum model \cite{hu2025amplitude}. Unlike angle-encoding methods, where features are mapped to rotation angles, AE-CQTL encodes the feature vector into the amplitudes of a quantum state. This allows a $2^n$-dimensional normalized vector to be represented using $n$ qubits.

In our benchmark implementation, the frozen ResNet18 backbone produces a $512$-dimensional feature vector. Since $512=2^9$, the AE-CQTL configuration uses $9$ qubits to amplitude-encode the backbone representation. The quantum layer follows a TLQNN-style design with repeated variational blocks. Each block applies Hadamard gates, trainable rotation gates, and cyclic CNOT entanglement across the qubits. The final quantum representation is obtained by measuring Pauli-$Z$ expectation values from all qubits, which are then passed to a classical linear classifier.

Within the benchmark, AE-CQTL represents the amplitude-encoding family. It is also used for circuit-depth analysis because its TLQNN-style multi-layer ansatz allows the effect of increasing quantum depth to be studied directly.

\subsection{PVCQTL}

PVCQTL is included to evaluate post-variational strategies for classical-quantum transfer learning \cite{yogaraj2025post}. Instead of relying only on standard variational quantum layers and single-qubit Pauli-$Z$ measurements, PVCQTL introduces structured observable measurements that can extract multi-qubit correlation features from the quantum state. In this benchmark, two PVCQTL variants are considered: a modified observable-based version and a variational-post-variational version.

\subsubsection{Modified PVCQTL}

In the modified PVCQTL variant, the frozen ResNet18 backbone extracts a $512$-dimensional feature vector. A classical pre-network maps this representation to the number of qubits, and the resulting vector is scaled using a $\tanh$ activation before being encoded through $R_Y$ rotations. The quantum circuit uses a controlled-$Z$ entanglement pattern. Instead of measuring only standard single-qubit Pauli-$Z$ expectation values, this variant evaluates structured Pauli observables defined by a locality parameter. These observables allow the model to capture multi-qubit correlations before passing the resulting features to a classical post-network for classification.

\subsubsection{Variational PVCQTL}

The variational PVCQTL variant extends the modified observable-based design by adding a shallow trainable ansatz to the quantum circuit. After the same feature extraction, projection, and $R_Y$ encoding steps, the circuit applies a trainable variational layer composed of $R_X$, $R_Y$, and $R_Z$ rotations on each qubit, followed by an additional controlled-$Z$ entanglement layer. As in the modified version, the output is obtained through structured Pauli observable measurements, and the resulting quantum features are processed by a classical post-network.

Within the benchmark, PVCQTL represents a measurement-design-focused transfer strategy. Its main difference from dressed QTL and QPIE-inspired QTL is that it emphasizes structured observable construction and post-variational feature extraction rather than standard Pauli-$Z$ readout alone.
\begin{table}[t]
\centering
\caption{Summary of the QTL techniques compared in the benchmark.}
\label{tab:qtl_methods_summary}
\scriptsize
\setlength{\tabcolsep}{3pt}
\renewcommand{\arraystretch}{1.05}

\begin{tabularx}{\columnwidth}{
@{}
>{\raggedright\arraybackslash}p{0.16\columnwidth}
>{\raggedright\arraybackslash}p{0.48\columnwidth}
>{\raggedright\arraybackslash}X
@{}}
\toprule
\textbf{Method} & \textbf{Design} & \textbf{Benchmark Role} \\
\midrule

DQN-QTL
& Frozen CNN features are reduced and encoded using $R_Y$ rotations. The classifier uses trainable rotation layers and cyclic CNOT entanglement.
& Compact angle-encoded QTL baseline. \\

\addlinespace[2pt]

QPIE-inspired QTL
& Reduced features are encoded through $R_X$, $R_Y$, and $R_Z$ rotations, followed by variational layers with cyclic CNOT entanglement.
& Examines whether richer input encoding improves feature transfer. \\

\addlinespace[2pt]

AE-CQTL
& The $512$-dimensional backbone feature vector is amplitude encoded and processed by a multilayer TLQNN-style variational circuit.
& Examines high-dimensional encoding and circuit-depth scaling. \\

\addlinespace[2pt]

PVCQTL
& Reduced features are encoded using $R_Y$ rotations and processed by a CZ-based circuit with structured Pauli measurements.
& Examines structured observables and post-variational readout. \\

\addlinespace[2pt]

ED-QTL
& A quantum student receives reduced transferred features and is optimized using a teacher-guided distillation loss.
& Examines the effect of teacher guidance on quantum training. \\

\bottomrule
\end{tabularx}
\end{table}
\subsection{ED-QTL}

ED-QTL is included as a knowledge-transfer-based strategy \cite{hasan2023bridging}. Instead of training the quantum student only from ground-truth labels, the student is guided by soft predictions from classical teacher models. This allows the quantum model to learn from the decision structure of stronger classical networks while keeping the quantum student compact.

In our benchmark implementation, three teacher models are used: ResNet18, ResNet34, and DenseNet121. The teacher models are trained independently, and their output logits are aggregated by averaging before computing the distillation loss. During teacher training, the backbone networks are frozen and only the classifier heads are optimized. After teacher training, the teacher parameters remain fixed, and the quantum student is trained using a distillation objective that combines supervised learning with teacher-guided soft targets.

Within the benchmark, ensemble-distillation QTL represents a classical-to-quantum knowledge transfer strategy. It differs from the other methods because the quantum student is trained not only from hard labels, but also from the predictive behavior of the teacher ensemble.
Table~\ref{tab:qtl_methods_summary} summarizes the main design differences among the evaluated QTL techniques and their roles in the benchmark.

\begin{table*}[bt]
\centering
\caption{Main benchmark comparison across datasets. Results are shown as mean $\pm$ standard deviation across random seeds.}
\label{tab:benchmark_main}
\footnotesize
\begin{tabularx}{\textwidth}{llccccccc}
\toprule
 & Method & Qubits & Depth & Accuracy (\%) & Precision & Recall & F1-score & ROC-AUC \\
\midrule
\multirow{13}{*}{\rotatebox{90}{Fashion-MNIST}}
& DQN-QTL (q=4) & 4 & 2 & 81.16 $\pm$ 0.00 & 0.80 $\pm$ 0.00 & 0.81 $\pm$ 0.00 & 0.80 $\pm$ 0.00 & 0.97 $\pm$ 0.00 \\
& DQN-QTL (q=6)         & 6 & 2 & 88.22 $\pm$ 0.68 & 0.88 $\pm$ 0.01 & 0.88 $\pm$ 0.01 & 0.88 $\pm$ 0.01 & 0.99 $\pm$ 0.00 \\
& QPIE-inspired QTL (q=4, $\sim$ 40 epochs) & 4 & 2 & 84.98 $\pm$ 0.00 & 0.85 $\pm$ 0.00 & 0.85 $\pm$ 0.00 & 0.85 $\pm$ 0.00 & 0.98 $\pm$ 0.00 \\
& QPIE-inspired QTL (q=6) & 6 & 2 & \textbf{88.28 $\pm$ 0.16} & \textbf{0.88 $\pm$ 0.00} & \textbf{0.88 $\pm$ 0.00} & \textbf{0.88 $\pm$ 0.00} & \textbf{0.99 $\pm$ 0.00} \\
& AE-CQTL (d=2) & 9 & 2 & 65.30 $\pm$ 0.26 & 0.64 $\pm$ 0.01 & 0.65 $\pm$ 0.01 & 0.65 $\pm$ 0.00 & 0.94 $\pm$ 0.00 \\
& AE-CQTL (d=4) & 9 & 4 & 71.78 $\pm$ 0.44 & 0.71 $\pm$ 0.01 & 0.72 $\pm$ 0.00 & 0.71 $\pm$ 0.01 & 0.96 $\pm$ 0.00 \\
& AE-CQTL (d=6) & 9 & 6 & 74.98 $\pm$ 0.48 & 0.75 $\pm$ 0.00 & 0.75 $\pm$ 0.00 & 0.75 $\pm$ 0.00 & 0.97 $\pm$ 0.00 \\
& PVCQTL (Modified, $\sim$ 40 epochs) & 4 & 1 & 87.35 $\pm$ 0.27 & 0.88 $\pm$ 0.00 & 0.87 $\pm$ 0.00 & 0.87 $\pm$ 0.00 & 0.99 $\pm$ 0.00 \\
& PVCQTL (Modified, $\sim$ 40 epochs) & 6 & 1 & 87.15 $\pm$ 0.57 & 0.87 $\pm$ 0.00 & 0.87 $\pm$ 0.01 & 0.87 $\pm$ 0.01 & 0.99 $\pm$ 0.00 \\
& PVCQTL (Variational, $<$ 40 epochs) & 4 & 1 & 88.24 $\pm$ 0.24 & 0.88 $\pm$ 0.00 & 0.88 $\pm$ 0.00 & 0.87 $\pm$ 0.01 & 0.99 $\pm$ 0.00 \\
& PVCQTL (Variational, $<$ 40 epochs) & 6 & 1 & 87.49 $\pm$ 0.11 & 0.88 $\pm$ 0.00 & 0.87 $\pm$ 0.00 & 0.88 $\pm$ 0.00 & 0.99 $\pm$ 0.00 \\
& ED-QTL (q=4, $\sim$ 40 epochs) & 4 & 2 & 85.45 $\pm$ 0.00 & 0.85 $\pm$ 0.00 & 0.85 $\pm$ 0.00 & 0.85 $\pm$ 0.00 & 0.98 $\pm$ 0.00 \\
& ED-QTL (q=6) & 6 & 2 & 85.88 $\pm$ 0.39 & 0.86 $\pm$ 0.00 & 0.86 $\pm$ 0.00 & 0.86 $\pm$ 0.00 & 0.99 $\pm$ 0.00 \\
\midrule
\multirow{13}{*}{\rotatebox{90}{Ants vs. Bees}} 
& DQN-QTL (q=4, $<$ 40 epochs) & 4 & 2 & 93.46 $\pm$ 0.00 & 0.93 $\pm$ 0.00 & 0.94 $\pm$ 0.00 & 0.93 $\pm$ 0.00 & 0.97 $\pm$ 0.00 \\
& DQN-QTL (q=6) & 6 & 2 & 91.50 $\pm$ 0.00 & 0.92 $\pm$ 0.00 & 0.92 $\pm$ 0.00 & 0.91 $\pm$ 0.00 & 0.97 $\pm$ 0.00 \\
& QPIE-inspired QTL (q=4, $<$ 40 epochs) & 4 & 2 & 90.85 $\pm$ 0.00 & 0.91 $\pm$ 0.00 & 0.91 $\pm$ 0.00 & 0.91 $\pm$ 0.00 & 0.97 $\pm$ 0.00 \\
& QPIE-inspired QTL (q=6, $<$ 40 epochs) & 6 & 2 & 90.20 $\pm$ 0.00 & 0.90 $\pm$ 0.00 & 0.91 $\pm$ 0.00 & 0.90 $\pm$ 0.00 & 0.96 $\pm$ 0.00 \\
& AE-CQTL (d=2) & 9 & 2 & 94.12 $\pm$ 0.66 & 0.94 $\pm$ 0.01 & 0.94 $\pm$ 0.01 & 0.94 $\pm$ 0.01 & 0.99 $\pm$ 0.00 \\
& AE-CQTL (d=4) & 9 & 4 & 94.77 $\pm$ 0.65 & 0.95 $\pm$ 0.01 & 0.95 $\pm$ 0.01 & 0.95 $\pm$ 0.01 & 0.99 $\pm$ 0.00 \\
& AE-CQTL (d=6) & 9 & 6 & \textbf{94.99 $\pm$ 0.38} & \textbf{0.95 $\pm$ 0.00} & \textbf{0.95 $\pm$ 0.00} & \textbf{0.95 $\pm$ 0.00} & \textbf{0.99 $\pm$ 0.00} \\
& PVCQTL (Modified, $\sim$ 40 epochs) & 4 & 2 & 86.71 $\pm$ 5.64 & 0.88 $\pm$ 0.04 & 0.88 $\pm$ 0.05 & 0.87 $\pm$ 0.06 & 0.95 $\pm$ 0.04 \\
& PVCQTL (Modified) & 6 & 2 & 93.03 $\pm$ 1.64 & 0.93 $\pm$ 0.02 & 0.93 $\pm$ 0.02 & 0.93 $\pm$ 0.02 & 0.97 $\pm$ 0.01 \\
& PVCQTL (Variational) & 4 & 2 & 87.58 $\pm$ 4.07 & 0.88 $\pm$ 0.03 & 0.88 $\pm$ 0.04 & 0.88 $\pm$ 0.04 & 0.96 $\pm$ 0.01 \\
& PVCQTL (Variational) & 6 & 2 & 91.50 $\pm$ 2.62 & 0.92 $\pm$ 0.02 & 0.92 $\pm$ 0.02 & 0.92 $\pm$ 0.03 & 0.97 $\pm$ 0.00 \\
& ED-QTL (q=4) & 4 & 2 & 71.24 $\pm$ 5.03 & 0.71 $\pm$ 0.05 & 0.71 $\pm$ 0.05 & 0.71 $\pm$ 0.05 & 0.79 $\pm$ 0.03 \\
& ED-QTL (q=6) & 6 & 2 & 74.51 $\pm$ 3.28 & 0.75 $\pm$ 0.03 & 0.75 $\pm$ 0.03 & 0.74 $\pm$ 0.03 & 0.81 $\pm$ 0.03 \\
\midrule
\multirow{5}{*}{\rotatebox{90}{CIFAR-10}}
& DQN-QTL (q=4, $\sim$ 40 epochs) & 4 & 2 & 71.79 $\pm$ 0.00 & 0.72 $\pm$ 0.00 & 0.72 $\pm$ 0.00 & 0.71 $\pm$ 0.00 & 0.95 $\pm$ 0.00 \\
& DQN-QTL (q=6, $<$ 40 epochs) & 6 & 2 & \textbf{77.19 $\pm$ 0.00} & \textbf{0.78 $\pm$ 0.00} & \textbf{0.77 $\pm$ 0.00} & \textbf{0.77 $\pm$ 0.00} & \textbf{0.97 $\pm$ 0.00} \\
& AE-CQTL (d=2) & 9 & 2 & 49.91 $\pm$ 0.00 & 0.50 $\pm$ 0.00 & 0.50 $\pm$ 0.00 & 0.49 $\pm$ 0.00 & 0.89 $\pm$ 0.00 \\
& AE-CQTL (d=4) & 9 & 4 & 55.35 $\pm$ 0.00 & 0.55 $\pm$ 0.00 & 0.55 $\pm$ 0.00 & 0.55 $\pm$ 0.00 & 0.91 $\pm$ 0.00 \\
& AE-CQTL (d=6, $<$ 40 epochs) & 9 & 6 & 56.59 $\pm$ 0.00 & 0.56 $\pm$ 0.00 & 0.57 $\pm$ 0.00 & 0.56 $\pm$ 0.00 & 0.92 $\pm$ 0.00 \\
\bottomrule
\end{tabularx}%

\end{table*}
\section{Results and Discussion}
\subsection{Experimental Setup}
\label{sec:experimental_setup}

Fashion-MNIST and Hymenoptera Ants vs.\ Bees were used as the main benchmark datasets, while CIFAR-10 was evaluated for selected configurations. Fashion-MNIST and CIFAR-10 used their standard train/test splits, and the Hymenoptera validation set was used for held-out evaluation. All images were resized to \(224\times224\) and normalized using ImageNet statistics. Fashion-MNIST images were converted to three channels, while Hymenoptera training images used random horizontal flips and small rotations.

All QTL models used a frozen ResNet18 backbone that produced \(512\)-dimensional feature vectors. These features were projected to a quantum-compatible size or amplitude encoded, depending on the method. AE-CQTL used \(9\) qubits because \(512=2^9\). ED-QTL used ResNet18, ResNet34, and DenseNet121 teachers with averaged logits and was evaluated on Fashion-MNIST and Hymenoptera.

Most methods were evaluated with \(4\) and \(6\) qubits at depth \(2\), while AE-CQTL used depths \(d\in\{2,4,6\}\). Models were trained for up to \(80\) epochs using Adam, a learning rate of \(10^{-3}\), and batch size \(32\). Standard QTL models used cross-entropy loss. ED-QTL combined cross-entropy with Kullback--Leibler divergence using \(T=2.0\) and \(\alpha=0.4\). Weight decay of \(10^{-4}\) was applied to DQN-QTL, QPIE-QTL, and ED-QTL, while StepLR with step size \(20\) and decay factor \(0.1\) was used for DQN-QTL and QPIE-QTL.

To reduce simulation cost, the \(4\)-qubit DQN-QTL and QPIE-QTL Fashion-MNIST experiments used a class-balanced subset of \(10{,}000\) samples. Other configurations used the full training set unless stated otherwise. Runs were stopped early when performance saturated or simulation cost became excessive, as marked in the result tables.
Experiments used seeds \(\{42,123,600\}\), and results are reported as mean \(\pm\) standard deviation. Models were implemented in PyTorch and PennyLane using the \texttt{default.qubit} simulator with analytic expectation values. Classical operations ran on NVIDIA A100 GPUs, while quantum circuit evaluations were executed on CPU.

\subsection{Benchmark Performance Across QTL Families}
Table~\ref{tab:benchmark_main} summarizes the QTL results. Performance varies by dataset and transfer design. On Fashion-MNIST, QPIE-QTL with \(6\) qubits achieves \(88.28\%\), closely followed by the \(4\)-qubit variational PVCQTL variant (PVCQTL-V) at \(88.24\%\) and \(6\)-qubit DQN-QTL at \(88.22\%\). These small differences indicate that cost and circuit design are more informative than accuracy alone. AE-CQTL improves with depth from \(65.30\%\) to \(74.98\%\), but remains below the leading configurations.

\begin{figure*}[b]
    \centering
    \includegraphics[width=1\linewidth]{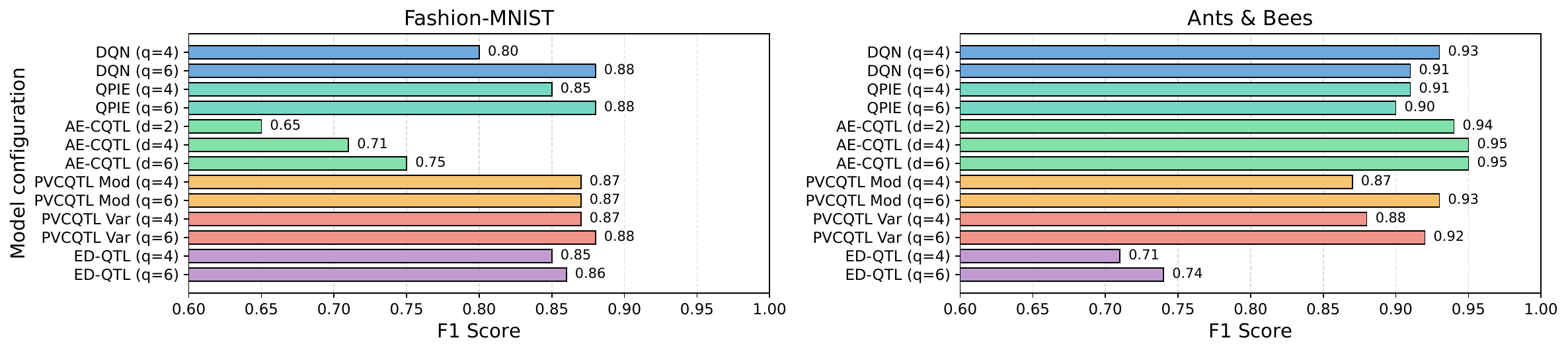}
    \caption{F1 score comparison across QTL configurations on Fashion-MNIST and Ants vs. Bees. Each bar represents one model variant, and colors indicate the corresponding QTL family. The side-by-side layout highlights dataset-dependent behavior: several DQN, QPIE, and PVCQTL variants reach closely matched F1 scores on Fashion-MNIST, while AE-CQTL provides the strongest F1 scores on Ants vs. Bees.} 
    \label{fig:main_f1_comparison}
\end{figure*}

On Ants vs.\ Bees, AE-CQTL reaches the highest accuracy, \(94.99\%\), with an F1-score of \(0.95\) at depth \(6\). DQN-QTL with \(4\) qubits reaches \(93.46\%\), and the \(6\)-qubit modified PVCQTL variant (PVCQTL-M) reaches \(93.03\%\). ED-QTL performs substantially lower, showing that teacher guidance does not consistently improve the quantum student.

For CIFAR-10, \(6\)-qubit DQN-QTL improves over its \(4\)-qubit version from \(71.79\%\) to \(77.19\%\). AE-CQTL also improves with depth from \(49.91\%\) to \(56.59\%\). Since only selected methods were evaluated, these results provide configuration-level evidence rather than a full cross-family ranking.

As shown in Fig.~\ref{fig:main_f1_comparison}, no QTL family dominates across all datasets. QPIE-QTL and PVCQTL are competitive on Fashion-MNIST, AE-CQTL is strongest on Ants vs.\ Bees, and DQN-QTL performs best among the tested CIFAR-10 configurations. This supports evaluating QTL methods by performance, cost, and architectural sensitivity rather than by a single accuracy result.

\subsection{Performance-Cost Trade-Offs\label{performance}}

Table~\ref{tab:complexity_efficiency} compares total parameters, quantum parameters, and training time across QTL configurations. The results show that similar accuracy can require very different computational costs.
\begin{table}[htpb]
\centering
\caption{Complexity and efficiency comparison across QTL configurations. Training time is reported in seconds.}
\label{tab:complexity_efficiency}
\scriptsize
\setlength{\tabcolsep}{2pt}
\renewcommand{\arraystretch}{1.05}

\resizebox{\columnwidth}{!}{%
\begin{tabular}{@{}llccrrr@{}}
\toprule
\textbf{Dataset} & \textbf{Method} & \textbf{Q} & \textbf{D} &
\textbf{Total Params} & \textbf{Q. Params} & \textbf{Time (s)} \\
\midrule

\multirow{13}{*}{\rotatebox{90}{Fashion-MNIST}}
& DQN-QTL ($q=4$)          & 4 & 2 & 11,666,254 & 24  & 4,905.51 \\
& DQN-QTL ($q=6$)          & 6 & 2 & 11,650,836 & 36  & 148,378.10 \\
& QPIE-inspired ($q=4$)    & 4 & 2 & 11,680,554 & 24  & 156,475.24 \\
& QPIE-inspired ($q=6$)    & 6 & 2 & 11,680,566 & 36  & 181,675.96 \\
& AE-CQTL ($d=2$)          & 9 & 2 & 11,250,441 & 81  & 2,599.71 \\
& AE-CQTL ($d=4$)          & 9 & 4 & 11,250,495 & 135 & 35,491.19 \\
& AE-CQTL ($d=6$)          & 9 & 6 & 11,250,549 & 189 & 50,006.08 \\
& PVCQTL (Mod., $q=4$)     & 4 & 1 & 11,178,574 & 16  & 352,485.60 \\
& PVCQTL (Mod., $q=6$)     & 6 & 1 & 11,178,590 & 24  & 329,580.45 \\
& PVCQTL (Var., $q=4$)     & 4 & 1 & 11,178,586 & 16  & 265,482.15 \\
& PVCQTL (Var., $q=6$)     & 6 & 1 & 11,178,602 & 24  & 441,753.12 \\
& ED-QTL ($q=4$)           & 4 & 2 & 11,244,274     & 24  & 334.18 \\
& ED-QTL ($q=6$)           & 6 & 2 & 11,250,538     & 36  & 2,849.37 \\
\midrule

\multirow{13}{*}{\rotatebox{90}{Ants vs. Bees}}
& DQN-QTL ($q=4$)          & 4 & 2 & 11,666,214 & 24  & 449.81 \\
& DQN-QTL ($q=6$)          & 6 & 2 & 11,650,780    & 36  & 1,954.97 \\
& QPIE-inspired ($q=4$)    & 4 & 2 & 11,680,514 & 24  & 695.64 \\
& QPIE-inspired ($q=6$)    & 6 & 2 & 11,680,510 & 36  & 1,160.20 \\
& AE-CQTL ($d=2$)          & 9 & 2 & 11,250,361 & 81  & 268.15 \\
& AE-CQTL ($d=4$)          & 9 & 4 & 11,250,415 & 135 & 350.12 \\
& AE-CQTL ($d=6$)          & 9 & 6 & 11,250,469 & 189 & 424.54 \\
& PVCQTL (Mod., $q=4$)     & 4 & 2 & 11,178,574     & 16  & 4,173.28 \\
& PVCQTL (Mod., $q=6$)     & 6 & 2 & 11,178,590     & 24  & 11,580.95 \\
& PVCQTL (Var., $q=4$)     & 4 & 2 & 11,178,586     & 16  & 4,600.36 \\
& PVCQTL (Var., $q=6$)     & 6 & 2 & 11,178,602   & 24  & 12,216.12 \\
& ED-QTL ($q=4$)           & 4 & 2 & 11,244,274     & 24  & 559.85 \\
& ED-QTL ($q=6$)           & 6 & 2 & 11,244,298     & 36  & 1,261.24 \\
\midrule

\multirow{5}{*}{\rotatebox{90}{CIFAR-10}}
& DQN-QTL ($q=4$)          & 4 & 2 & 11,666,254 & 24  & 11,677.97 \\
& DQN-QTL ($q=6$)          & 6 & 2 & 11,650,836 & 36  & 11,715.51 \\
& AE-CQTL ($d=2$)          & 9 & 2 & 11,250,441 & 81  & 20,037.36 \\
& AE-CQTL ($d=4$)          & 9 & 4 & 11,250,495 & 135 & 29,467.68 \\
& AE-CQTL ($d=6$)          & 9 & 6 & 11,250,549 & 189 & 518.54 \\
\bottomrule
\end{tabular}%
}
\end{table}
On Fashion-MNIST, QPIE-QTL, PVCQTL-V, and DQN-QTL achieve nearly identical leading accuracies but differ sharply in runtime. PVCQTL is the most costly, while ED-QTL is much faster but less accurate. AE-CQTL shows a clear depth--cost trade-off, with accuracy and runtime both increasing as depth grows.

On Ants vs.\ Bees, AE-CQTL at depth \(6\) provides the strongest observed balance, achieving the highest accuracy with far lower runtime than PVCQTL. DQN-QTL with \(4\) qubits is also competitive at low cost, whereas ED-QTL does not convert its moderate runtime into stronger predictive performance.

For CIFAR-10, DQN-QTL with \(6\) qubits performs best among the tested configurations. AE-CQTL improves with depth, but the depth-\(6\) runtime is not directly comparable because the run stopped before the full epoch budget.

\begin{figure*}[htpb]
    \centering
    \includegraphics[width=1\linewidth]{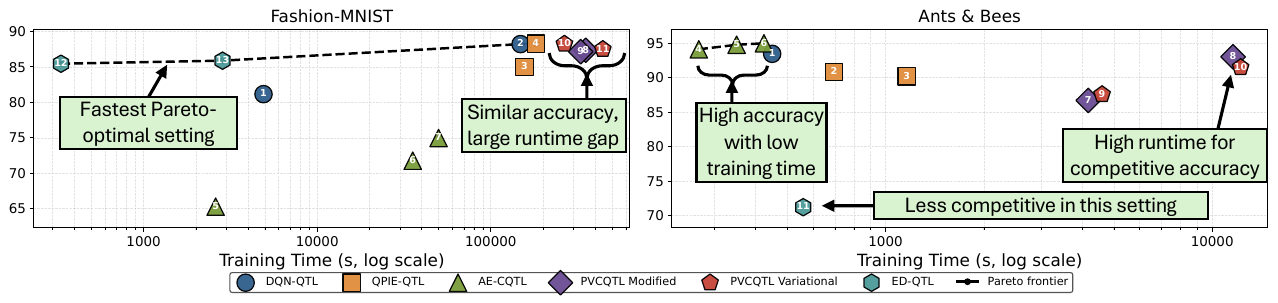}
    \caption{Performance-cost trade-off across QTL configurations on Fashion-MNIST and Ants vs. Bees. Each point represents one model configuration, with training time on the x-axis (log scale) and accuracy on the y-axis; numeric labels identify the corresponding variants, and the dashed line marks the Pareto frontier. Points closer to the upper-left region indicate a better balance between predictive performance and training cost.} 
    \label{fig:performance_cost_pareto}
\end{figure*}

This trend is illustrated in Fig.~\ref{fig:performance_cost_pareto}, where methods with similar accuracy exhibit substantially different training times. QTL evaluation should therefore consider predictive performance, circuit size, parameter count, and runtime jointly.

\subsection{Architectural Scaling Analysis}
\label{sec:architectural_scaling}

We examine the effect of circuit width and depth on QTL performance. Since both increase simulation cost, their gains should be interpreted with the trade-offs in Section~\ref{performance}.

\begin{figure*}[htpbt]
    \centering
    \includegraphics[width=\linewidth]{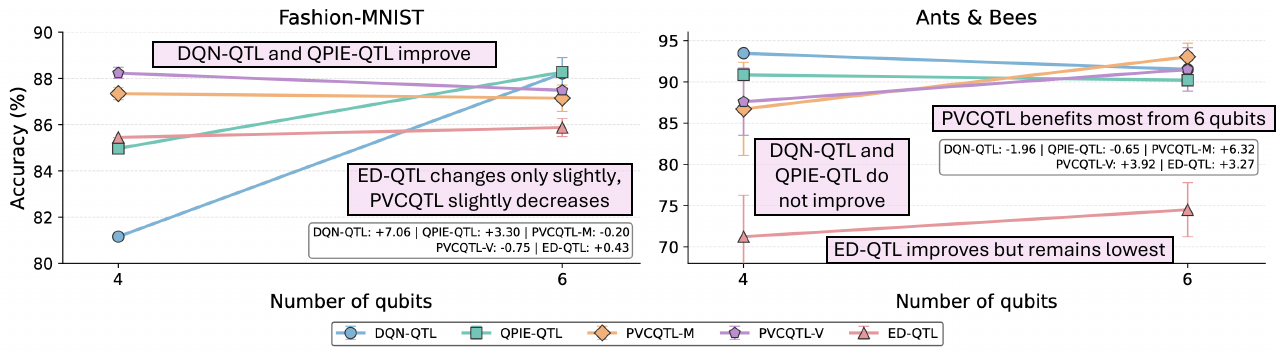}
    \caption{Qubit-scaling behavior of QTL configurations on the two main benchmark datasets. Increasing the number of qubits from $4$ to $6$ improves some methods on Fashion-MNIST, while the same trend is weaker or reversed on Hymenoptera Ants vs. Bees.}
    \label{fig:qubit_scaling}
\end{figure*}

Figure~\ref{fig:qubit_scaling} compares \(4\)- and \(6\)-qubit configurations. On Fashion-MNIST, DQN-QTL improves from \(81.16\%\) to \(88.22\%\), while QPIE-QTL improves from \(84.98\%\) to \(88.28\%\). On Hymenoptera, the same increase reduces performance for both methods. PVCQTL also shows dataset-dependent behavior: the \(4\)-qubit variational model is slightly stronger on Fashion-MNIST, while the \(6\)-qubit modified model performs better on Hymenoptera. For CIFAR-10, DQN-QTL improves from \(71.79\%\) to \(77.19\%\), although this result applies only to the tested configurations.

Table~\ref{tab:depth_scaling} shows a clearer depth trend for AE-CQTL. Increasing depth from \(2\) to \(6\) improves accuracy by \(9.68\) percentage points on Fashion-MNIST, \(0.87\) points on Hymenoptera, and \(6.68\) points on CIFAR-10. The smaller Hymenoptera gain suggests that the pretrained features already provide strong class separation. The CIFAR-10 depth-\(6\) result should be interpreted within its reduced training budget.

These results show that wider or deeper circuits can improve performance, but the gain depends on the dataset and model design.

\begin{table}[htpb]
\centering
\caption{AE-CQTL depth-scaling results across datasets. Results are reported as mean $\pm$ standard deviation across available runs.}
\label{tab:depth_scaling}
\scriptsize
\setlength{\tabcolsep}{2.5pt}
\renewcommand{\arraystretch}{1.08}

\resizebox{\columnwidth}{!}{%
\begin{tabular}{@{}llccccc@{}}
\toprule
\textbf{Dataset} & \textbf{Method} & \textbf{Depth} &
\textbf{Accuracy (\%)} & \textbf{Precision} &
\textbf{Recall} & \textbf{F1-score} \\
\midrule

\multirow{3}{*}{Fashion-MNIST}
& AE-CQTL & 2
& $65.30 \pm 0.26$
& $0.64 \pm 0.01$
& $0.65 \pm 0.01$
& $0.65 \pm 0.00$ \\

& AE-CQTL & 4
& $71.78 \pm 0.44$
& $0.71 \pm 0.01$
& $0.72 \pm 0.00$
& $0.71 \pm 0.01$ \\

& AE-CQTL & 6
& $\mathbf{74.98 \pm 0.48}$
& $\mathbf{0.75 \pm 0.00}$
& $\mathbf{0.75 \pm 0.00}$
& $\mathbf{0.75 \pm 0.00}$ \\
\midrule

\multirow{3}{*}{Ants vs. Bees}
& AE-CQTL & 2
& $94.12 \pm 0.66$
& $0.94 \pm 0.01$
& $0.94 \pm 0.01$
& $0.94 \pm 0.01$ \\

& AE-CQTL & 4
& $94.77 \pm 0.65$
& $0.95 \pm 0.01$
& $0.95 \pm 0.01$
& $0.95 \pm 0.01$ \\

& AE-CQTL & 6
& $\mathbf{94.99 \pm 0.38}$
& $\mathbf{0.95 \pm 0.00}$
& $\mathbf{0.95 \pm 0.00}$
& $\mathbf{0.95 \pm 0.00}$ \\
\midrule

\multirow{3}{*}{CIFAR-10}
& AE-CQTL & 2
& $49.91 \pm 0.00$
& $0.50 \pm 0.00$
& $0.50 \pm 0.00$
& $0.49 \pm 0.00$ \\

& AE-CQTL & 4
& $55.35 \pm 0.00$
& $0.55 \pm 0.00$
& $0.55 \pm 0.00$
& $0.55 \pm 0.00$ \\

& AE-CQTL & 6
& $\mathbf{56.59 \pm 0.00}$
& $\mathbf{0.56 \pm 0.00}$
& $\mathbf{0.57 \pm 0.00}$
& $\mathbf{0.56 \pm 0.00}$ \\
\bottomrule
\end{tabular}%
}
\end{table}

\subsection{Classical Baselines}
\label{sec:classical_baselines}

Although the main scope of this study is the controlled comparison of QTL techniques, classical heads are included to show how much predictive performance is already provided by the frozen pretrained backbone. Table~\ref{tab:classical_noncached_baselines} reports non-cached baselines using the same ImageNet-pretrained ResNet18 feature extractor with either a linear or multilayer perceptron (MLP) head. Cached results are omitted from the main paper because they produce nearly identical predictive metrics and mainly reflect implementation-level runtime differences.

\begin{table*}[t]
\centering
\caption{End-to-end non-cached classical baselines using a frozen ImageNet-pretrained ResNet18 backbone. Results are reported as mean \(\pm\) standard deviation across seeds \(42\), \(123\), and \(600\). Precision, recall, and F1-score are weighted averages, while ROC-AUC is computed using weighted one-vs.-rest averaging. Total parameter counts include the frozen backbone and the trainable classification head.}
\label{tab:classical_noncached_baselines}
\resizebox{\textwidth}{!}{%
\begin{tabular}{lllrcccccc}
\toprule
Dataset & Training Set & Head & Total Parameters &
Accuracy (\%) & Precision & Recall &
ROC-AUC & F1-score & Train Time (s) \\
\midrule

Ants vs.\ Bees
& Full
& Linear
& 11,177,538
& \(94.55 \pm 0.75\)
& \(0.9468 \pm 0.0072\)
& \(0.9455 \pm 0.0075\)
& \(0.9819 \pm 0.0010\)
& \(0.9456 \pm 0.0075\)
& \(252.22 \pm 114.64\) \\

Ants vs.\ Bees
& Full
& MLP
& 11,242,434
& \(94.55 \pm 0.75\)
& \(0.9468 \pm 0.0072\)
& \(0.9455 \pm 0.0075\)
& \(0.9785 \pm 0.0019\)
& \(0.9456 \pm 0.0075\)
& \(185.47 \pm 0.48\) \\

\midrule

Fashion-MNIST
& 10K
& Linear
& 11,181,642
& \(87.21 \pm 0.23\)
& \(0.8734 \pm 0.0037\)
& \(0.8721 \pm 0.0023\)
& \(0.9888 \pm 0.0002\)
& \(0.8719 \pm 0.0032\)
& \(274.79 \pm 2.29\) \\

Fashion-MNIST
& 10K
& MLP
& 11,243,466
& \(87.04 \pm 0.42\)
& \(0.8709 \pm 0.0036\)
& \(0.8704 \pm 0.0042\)
& \(0.9893 \pm 0.0004\)
& \(0.8701 \pm 0.0041\)
& \(271.83 \pm 1.72\) \\

\midrule

Fashion-MNIST
& Full
& Linear
& 11,181,642
& \(89.08 \pm 0.11\)
& \(0.8909 \pm 0.0013\)
& \(0.8908 \pm 0.0011\)
& \(0.9916 \pm 0.0001\)
& \(0.8905 \pm 0.0012\)
& \(1727.76 \pm 1.63\) \\

Fashion-MNIST
& Full
& MLP
& 11,243,466
& \(89.98 \pm 0.43\)
& \(0.9013 \pm 0.0026\)
& \(0.8998 \pm 0.0043\)
& \(0.9932 \pm 0.0002\)
& \(0.8999 \pm 0.0040\)
& \(1746.46 \pm 0.67\) \\

\midrule

CIFAR-10
& Full
& Linear
& 11,181,642
& \(87.21 \pm 0.06\)
& \(0.8724 \pm 0.0004\)
& \(0.8721 \pm 0.0006\)
& \(0.9907 \pm 0.0000\)
& \(0.8718 \pm 0.0004\)
& \(1022.87 \pm 9.19\) \\

CIFAR-10
& Full
& MLP
& 11,243,466
& \(87.40 \pm 0.42\)
& \(0.8751 \pm 0.0037\)
& \(0.8740 \pm 0.0042\)
& \(0.9915 \pm 0.0003\)
& \(0.8733 \pm 0.0045\)
& \(981.47 \pm 14.67\) \\

\bottomrule
\end{tabular}%
}
\end{table*}

The classical heads outperform the strongest evaluated QTL configurations on Fashion-MNIST and CIFAR-10. On full Fashion-MNIST, the MLP reaches \(89.98 \pm 0.43\%\), compared with \(88.28 \pm 0.16\%\) for \(6\)-qubit QPIE-QTL. On CIFAR-10, the MLP reaches \(87.40 \pm 0.42\%\), compared with \(77.19\%\) for \(6\)-qubit DQN-QTL. On Ants vs.\ Bees, AE-CQTL reaches \(94.99 \pm 0.38\%\), while both classical heads obtain \(94.55 \pm 0.75\%\); this small numerical difference does not establish a reliable advantage.
These results confirm that the pretrained backbone provides much of the predictive signal. The benchmark therefore evaluates how effectively different QTL designs preserve and transform pretrained features under controlled encoding, circuit, measurement, and resource settings, rather than claiming consistent quantum superiority.

\subsection{Benchmark Scope: Clarifying Design Choices}
\label{sec:benchmark_scope}

To clarify the benchmark scope and interpretation, the main design choices are summarized below.

\textbf{Why were Fashion-MNIST and Hymenoptera selected?}
They represent two distinct settings: structured grayscale classification and low-data natural-image transfer.

\textbf{Why is CIFAR-10 treated separately?}
Only selected QTL families were evaluated on CIFAR-10. Its results therefore provide configuration-level evidence, while the main cross-family conclusions rely on Fashion-MNIST and Hymenoptera.

\textbf{Why not include more datasets?}
The benchmark prioritizes controlled comparison across methods, seeds, qubit counts, and circuit depths. Broader dataset coverage would substantially increase simulation cost and is left for future work.

\textbf{Why were these QTL techniques selected?}
They represent distinct design directions: dressed quantum heads, multi-axis encoding, amplitude encoding, structured observable measurements, and teacher-guided transfer.

\textbf{Why not evaluate larger qubit counts?}
The study focuses mainly on \(4\)- and \(6\)-qubit settings to keep simulation feasible and reflect near-term limits. AE-CQTL uses \(9\) qubits because \(512=2^9\), matching the backbone feature dimension.

\textbf{Why use simulation instead of quantum hardware?}
Simulation provides a controlled and reproducible comparison without device-specific noise, connectivity limits, calibration drift, or shot variation.

\textbf{Why do some results show \(0.00\) standard deviation?}
A value of \(0.00\) occurs when the variation across repeated runs is smaller than the reported decimal precision after rounding. It should therefore be interpreted as very low observed variation, not exact zero variability.

\textbf{Why are some runtimes non-monotonic?}
Runtime also depends on completed epochs, stopping conditions, simulator behavior, and per-sample circuit evaluation. Partial runs are therefore not used to support strict efficiency claims.

\textbf{Why report cost with accuracy?}
QTL models must be assessed by predictive performance, circuit size, parameter count, and training cost rather than accuracy alone.

\textbf{Why are classical baselines included?}
The classical heads provide a reference for the predictive information already supplied by the frozen backbone. They contextualize the QTL results but are not the main focus of the benchmark, which compares design choices across QTL families.

\subsection{Final Insights}
\label{sec:final_insights}

The benchmark yields five main findings:

\begin{itemize}
\item \textbf{Performance is dataset-dependent.}
QPIE-QTL and PVCQTL are among the leading configurations on Fashion-MNIST, AE-CQTL achieves the highest reported performance on Hymenoptera, and DQN-QTL achieves the highest reported performance among the tested CIFAR-10 configurations.
    \item \textbf{More qubits do not always help.}
    Increasing from \(4\) to \(6\) qubits improves some Fashion-MNIST results but gives smaller or negative changes on Hymenoptera.

    \item \textbf{AE-CQTL shows the clearest depth trend.}
    Greater depth improves AE-CQTL on all three datasets, with the largest gains on Fashion-MNIST and CIFAR-10.

    \item \textbf{Accuracy and cost must be considered jointly.}
    Configurations with similar accuracy can differ greatly in runtime and quantum parameters.

\item \textbf{Model choice should follow the target constraint.}
DQN-QTL and QPIE-QTL are suitable for small qubit budgets, AE-CQTL for depth-scaling studies, PVCQTL for structured measurements, and ED-QTL for settings where distillation can be carefully tuned.
\end{itemize}

\section{Conclusion}
\label{sec:conclusion}

This work introduced a controlled benchmark for evaluating QTL methods under shared datasets, preprocessing, frozen-backbone settings, training conditions, and reporting metrics. By comparing representative QTL families within a common transfer-learning pipeline, the study provides a consistent basis for evaluating predictive performance, circuit design, and computational cost.
The results show that QTL performance depends on the dataset and configuration, with no single family dominating across all settings. Increasing qubit count or circuit depth improves some models but gives limited or negative gains in others. Classical baselines further show that the frozen pretrained backbone supplies much of the predictive signal, while the QTL comparison reveals how encoding, circuit, and measurement choices affect the retention of transferred features.
These findings support resource-aware QTL evaluation based on accuracy, circuit size, parameter count, and training time. Future work will extend the benchmark to additional datasets, broader QTL architectures, noisy simulations, real quantum hardware, and shot-based evaluation.

\small{
\section*{Acknowledgment}
 This work was supported in part by the NYUAD Center for Quantum and Topological Systems (CQTS), funded by Tamkeen under the NYUAD Research Institute grant CG008, and the Center for Cyber Security (CCS), funded by Tamkeen under the NYUAD Research Institute Award G1104. This research was carried out on the High Performance Computing resources at New York University Abu Dhabi.}
\small{\section*{Code availability} The source code and experiment configurations are available at \url{https://github.com/eBrain4Everyone/QTL-Benchmark}.}

\bibliographystyle{IEEEtran}

{\setstretch{0.88}

{\scriptsize
\bibliography{refs}}}

\end{document}